\documentclass[conference]{IEEEtran}

\usepackage[cmex10]{amsmath}
\usepackage[pdftex]{graphicx}
\usepackage{latexsym}
\usepackage{amssymb}
\usepackage{bm}
\usepackage{psfrag}
\usepackage{algorithm}
\usepackage{algorithmic}

\newtheorem{theorem}{Theorem}
\newtheorem{lemma}{Lemma}

\newtheorem{remark}{Remark}

\newtheorem{example}{Example}

\newcommand{\argmax}{\operatornamewithlimits{argmax}}

\newcommand{\bs}{\boldsymbol}

\begin{document}

\sloppy

\title{
An Improvement of Non-binary Code Correcting Single $b$-Burst of Insertions or Deletions
}

\author{
\IEEEauthorblockN{Toyohiko Saeki and Takayuki Nozaki}
\IEEEauthorblockA{
Dept. of Informatics, Yamaguchi University, JAPAN\\
Email: \{g012vb,tnozaki\}@yamaguchi-u.ac.jp
}
}
%% Create the title:
\maketitle

%% Abstract: 
\begin{abstract}
This paper constructs a non-binary code correcting a single $b$-burst of insertions or deletions with a large cardinality.
This paper also proposes a decoding algorithm of this code and evaluates a lower bound of the cardinality of this code.
Moreover, we evaluate an asymptotic upper bound on the cardinality of codes which correct a single burst of insertions or deletions.
\end{abstract}

\section{Introduction}
In communication and storage systems, several symbols in a sequence are inserted or deleted for the synchronization errors.
Levenshtein \cite{levenshtein1966binary} proved that VT codes (constructed by Varshamov and Tenengolts \cite{varshamov1965code} for error correction on the Z-channel) correct a single insertion or deletion.
This code had been extended to non-binary single insertion or deletion \cite{tenengolts1984nonbinary} and to two adjacent insertion or deletion \cite{levenshtein1967asymptotically}.
This code had been also extended to a binary \cite{helberg2002multiple} and a non-binary multiple insertion or deletion correcting code \cite{paluncic2011note}.

Cheng et al.\ \cite{cheng2014codes} constructed a binary $b$-burst insertion or deletion correcting code, which corrects any consecutive insertion or deletion of length $b$.
Schoeny et al.\ \cite{schoeny2017codes} improved this construction and showed that the resulting code has larger cardinality than the code constructed by Cheng et al.
These constructions have been extended to permutation code \cite{chee2015permutation,chee2017permutation}.
Nowadays, Schoeny et al.\ \cite{schoeny2017novel} gives a non-binary $b$-burst insertion or deletion correcting code.

In this paper, we construct a non-binary $b$-burst insertion or deletion correcting code with a larger cardinality.
The key idea of the paper is to investigate the correcting capability of the non-binary shifted VT code, which is a component of non-binary $b$-burst insertion or deletion correcting codes.
We also derive a lower bound of the number of codewords of the constructed non-binary $b$-burst insertion or deletion correcting code.
Moreover, we show an asymptotic upper bound of the cardinality of the best non-binary $b$-burst insertion or deletion correcting code.

\section{Preliminaries And Previous Works}
This section briefly introduces previous works, i.e, insertion/deletion\footnote{Section \ref{ssec:not} will give the details of definition of the notation ``insertion/deletion''.} codes given in \cite{varshamov1965code,tenengolts1984nonbinary,cheng2014codes,schoeny2017codes,schoeny2017novel}.
We use notations given in this section throughout the paper.

\subsection{Notation and Definition \label{ssec:not}}
For integers $i,j$, define $[i,j] := \{k\in\mathbb{Z} \mid i \le k \le j\}$ and $[i] := [0,i-1]$, where $\mathbb{Z}$ stands the set of integers.
For a sequence ${\bs x} = (x_1,x_2, \dots, x_{n}) \in [q]^n$, we denote the subsequence of ${\bs x}$ whose $s$-th symbol is deleted, by ${\bs x}_{\neg s}$, i.e, ${\bs x}_{\neg s} = (x_1,x_2,\dots, x_{s-1}, x_{s+1},\dots, x_{n})$.
In this case, we say that a single deletion has occurred in ${\bs x}$.
If ${\bs y}$ is an output of the single insertion channel with an input ${\bs x}$, there exists $i$ such that ${\bs y}_{\neg i} = {\bs x}$.
For a sequence ${\bs x} \in [q]^n$, a symbol $\lambda\in[q]$, and an integer $s \in [1,n+1]$, we denote ${\bs x}_{\vdash (s,\lambda)} = (x_1,x_2,\dots, x_{s-1},\lambda,x_s,\dots, x_n)$.

%% run
A {\it run} of length $r$ of a sequence ${\bs x}$ is a subsequence of ${\bs x}$ such that $x_i = x_{i+1} = \cdots = x_{i+r-1}$, $x_{i-1}\neq x_i$ (for $i> 1$), and $x_{i+r-1}\neq x_{i+r}$ (for $i+r\le n$).

\begin{remark} \label{rem:1}
For a sequence ${\bs x} = (1,0,0,1,1,1)$, $(x_2,x_3)$ is a run of length 2 and we have
\begin{align*}
% &{\bs x}_{\neg 1} = (0,0,1,1,1), \\
 &{\bs x}_{\neg 2} = {\bs x}_{\neg 3} = (1,0,1,1,1).
\end{align*}
From this, we see that we receive the same subsequences if a symbol in the same run is deleted under the single deletion channel.
In other words, in the single deletion channel, even if one can correct a deletion, one cannot detect which symbol in a run is deleted.

Similarly, we get
\begin{align*}
 {\bs x}_{\vdash (2,0)} = {\bs x}_{\vdash (3,0)} = {\bs x}_{\vdash (4,0)} = (1,0,0,0,1,1,1).
\end{align*}
Hence, in the single insertion channel, we receive the same sequence if the symbol $\lambda$ is inserted into a run of $\lambda$.
\end{remark}

We refer to exactly $b$ consecutive deletions as a single $b$-burst deletion.
We define ${\bs x}_{\neg [i+1,i+b]} := (x_1,x_2,\dots,\allowbreak x_{i},\allowbreak x_{i+b+1},x_{i+b+2},\dots,\allowbreak x_{n})$.
In words, when the $b$ consecutive, namely from $i$-th to $(i+b-1)$-th, symbols of ${\bs x}$ are deleted, we denote it, by ${\bs x}_{\neg [i,i+b-1]}$.
If ${\bs y}$ is an output of the single $b$-insertion channel with an input ${\bs x}$, there exists an integer $i$ such that ${\bs y}_{\neg [i,i+b-1]} = {\bs x}$.

A code which corrects single $b$-burst deletions (resp.\ insertions) is called a {\it single $b$-burst deletion (resp.\ insertion) correcting code}.
A code is {\it $b$-burst insertion/deletion correcting} if it corrects single $b$-burst insertions or single $b$-burst deletions.
Similarly, we define the terms: {\it single deletion correcting code}, {\it single insertion correcting code}, and {\it single insertion/deletion correcting code}.

The following theorem given in \cite{schoeny2017codes} shows a relationship between single $b$-burst deletion correcting codes and single $b$-burst insertion correcting codes.
\begin{theorem}{ \cite[Theorem 1]{schoeny2017codes} } 
 A code is a $b$-burst deletion correcting code if and only if it is a $b$-burst insertion correcting code.
\end{theorem}
This theorem holds for not only binary case but also non-binary case.
Hence, when we prove a code is a $b$-burst insertion/deletion correcting code, we only need to prove it is a $b$-burst deletion correcting code.

\subsection{Single Insertion/Deletion Correcting Code \label{ssec:VT}}
The VT code is a single insertion/deletion correcting code.
The VT code is defined by the code length $n$ and $a \in [n+1]$ as follows:
\begin{equation*}
 {\rm VT}_{a}(n) 
  =
 \bigl\{ {\bs x} \in [2]^n \mid 
 {\textstyle \sum_{i=1}^n i x_i } \equiv a \pmod {n+1} \bigr\}.
\end{equation*}

Let $\mathbb{I}[P]$ be the indicator function, which equals 1 if the proposition $P$ is true and equals 0 otherwise.
A mapping $\sigma$ of a $q$-ary sequence $(x_1,x_2,\dots, x_{n})\in[q]^n$ to a binary sequence $(u_1,u_2,\dots,u_{n-1})\in [2]^{n-1}$ is defined by 
\begin{equation*}
 u_i = \mathbb{I}[x_i < x_{i+1}].
\end{equation*}
We refer to the sequence ${\bs u} = \sigma ({\bs x})$ as the {\it ascent sequence} for ${\bs x}$.
The non-binary VT code is a non-binary single insertion/deletion correcting code defined by the code length $n$, $a\in[n]$ and $c\in[q]$ as follows:
\begin{align*}
  {q\rm{VT}}_{a, c}(n,q)
  = \bigl\{\bm{x} \in [q]^n
  \mid &~ {\textstyle\sum_{i=1}^{n}} x_i \equiv c \pmod q ,\\
  &~\sigma( \bm{x} ) \in {\rm{VT}}_{a} (n-1) \bigr\}.
\end{align*}

\subsection{Binary Burst Insertion/Deletion Correcting Code}
This section briefly introduces the binary $b$-burst insertion/deletion correcting codes given in \cite{cheng2014codes,schoeny2017codes}.
Roughly speaking, those methods employ interleaving to construct the codes.

For simplicity, we assume that $n$ is divided by $b$.
The $b \times \frac{n}{b}$ matrix representation for a sequence ${\bs x}$ is given as
\begin{align}
  A_b (\bm{x}) =
  \begin{pmatrix}
    x_1 & x_{b+1} & \cdots & x_{n-b+1}\\
    x_2 & x_{b+2} & \cdots & x_{n-b+2}\\
    \vdots & \vdots & \ddots & \vdots\\
    x_b & x_{2b} & \cdots & x_{n}
  \end{pmatrix}. \label{eq:matrix_rep_x}
\end{align}
We denote the $i$-th row of this matrix, by $A_{b}(\bm{x})_i$.

\begin{example}
Consider the 3-burst deletion channel with an input ${\bs x} \in[2]^{12}$.
Assume that the output is ${\bs x}_{\neg [6,8]}$.
Then, these matrix representations are 
\begin{align*}
    A_3(\bm{x})
    &=
    \begin{pmatrix}
      x_1 & x_4 & x_7 & x_{10} \\
      x_2 & x_5 & x_8 & x_{11} \\
      x_3 & x_6 & x_9 & x_{12} \\
    \end{pmatrix}, \\
    A_3(\bm{x}_{\neg [6,8]})
    &=
    \begin{pmatrix}
      x_1 & x_4 & x_{10} \\
      x_2 & x_5 & x_{11} \\
      x_3 & x_9 & x_{12} \\
    \end{pmatrix}.
\end{align*}
From these, we see that $A_3({\bs x}_{\neg [6,8]})_i$ is a result of a single deletion to $A_3({\bs x})_i$.
Moreover, we see that when the $(1,i)$-th entry of $A_3({\bs x})$ is deleted, the $(j,i-1)$-th or $(j,i)$-th entry is deleted for $j\ge 2$.
\end{example}

From the above example, for recovering a single $b$-burst deletion, one needs to correct a single deletion for each row of the matrix representation.
Moreover, if one detects the position $i$ of deletion in the first row, one needs to correct a deletion for a given two adjacent positions $i-1,i$ in the other rows.

The code in \cite[Sect.III-C]{cheng2014codes} embeds a {\it marker} $(0,1,0,1,\dots)$ in the first row of the matrix representation to detect the deletion position and employs substitution-transposition codes \cite{abdel1998detecting} in the other rows to correct a single deletion for a given two adjacent positions.
Here, note that we are able to regard to the marker $(0,1,0,1,\dots)$ as a codeword of a VT code with maximum run length 1.

Schoeny et al.\ \cite{schoeny2017codes} improved the construction of this code.
The first row of the code in \cite{schoeny2017codes} is a run-length-limited VT code which is a VT code with maximum run length at most $r$.
From Remark \ref{rem:1}, one detects the interval of deletion position with the length at most $r$.
The other rows of the code are the {\it shifted-VT codes}, which correct a single deletion for a given $r+1$ adjacent positions.
Let $S_{n,q}(r)$ be the set of sequences in $[q]^n$ with maximum run length at most $r$.
Then, the run-length-limited VT code and shifted-VT (SVT) code are defined as
\begin{align*}
  &{\rm{RLL\mathchar`-VT}}_{a} (n,r) = {\rm{VT}}_{a} (n) \cap S_{n,2} (r), \\
  &{\rm{SVT}}_{d, e}(n,r) = \bigl\{\bm{x} \in [2]^n
  : \textstyle\sum_{i=1}^{n} ix_i \equiv d \pmod {r} ,\\
  &\hspace{40mm} \textstyle\sum_{i=1}^{n} x_i \equiv e \pmod{2} \bigr\},
\end{align*}
for $d\in [r]$ and $e\in [2]$.
By using those codes, the binary single $b$-burst correcting code is constructed as:
\begin{align*}
  C_{2,b} &= \{\bm{x} : A_b(\bm{x})_1 \in {\rm{RLL\mathchar`-VT}}_{a}(n/b,r),\\
  &\hspace{11mm} \forall i \in [2,b]~~ A_b(\bm{x})_i \in  {\rm{SVT}}_{d,e}(n/b,r+1)\}.
\end{align*}

\subsection{Decoding Algorithm for SVT codes}
In this section, we briefly introduce the decoding algorithm for the SVT codes.
The details of decoding algorithms are in \cite[Appendix C]{schoeny2017codes}.

Firstly, we consider the case of deletion correction.
Assume that we employ ${\rm SVT}_{d,e}(n,r)$.
Let ${\bs y} \in [2]^{n-1}$ be the received sequence.
Denote the first possible deletion position, by $k$.
The inputs of the deletion decoder are those, namely ${\bs y}$, $(d,e,n,r)$, and $k$.
We denote the estimated codeword, by ${\bs x}$.
Let $[s,t]$ be the interval of the run which contains the inserted symbol.
The outputs of the deletion decoder are a pair of the estimated codeword ${\bs x}$ and interval $[s,t]$.
We denote the deletion correcting algorithm for the SVT code, 
by ${\sf SVT{\mathchar`-}DC}({\bs y}, d,e,n,r, k) \to ({\bs x}, [s,t])$.
For example, we have ${\sf SVT{\mathchar`-}DC}(0011, 0,0,5,3, 2) \to (00011, [1,3])$.

Secondly, we consider the case of insertion correction.
Let ${\bs y} \in [2]^{n+1}$ be the received sequence.
Denote the first possible insertion position, by $k$.
We denote the estimated codeword, by ${\bs x}$.
Let $[s,t]$ be the interval of the run which contains the deleted symbol.
We denote the insertion correcting algorithm for the SVT code, 
by ${\sf SVT{\mathchar`-}IC}({\bs y}, d,e,n,r, k) \to ({\bs x}, [s,t])$.
For example, we have ${\sf SVT{\mathchar`-}IC}(000111, 0,0,5,3, 2) \to (00011, [4,5])$.
The notations ${\sf SVT{\mathchar`-}DC}$ and ${\sf SVT{\mathchar`-}IC}$ will be used in Section \ref{ssec:dec}.

\subsection{Non-binary Burst Insertion/Deletion Correcting Code}
This section introduces the non-binary $b$-burst insertion/deletion correcting code give in \cite{schoeny2017novel}.

By a straightforward construction, one obtains the non-binary $b$-burst insertion/deletion correcting code.
Similar to the construction of non-binary VT code, we employ the mapping $\sigma$ given in Sect.~\ref{ssec:VT}.
The non-binary run-length-limited VT code and the non-binary SVT code are defined as:
\begin{align*}
  &{{\rm{RLL}}\mathchar`-q{\rm{VT}}}_{a, c} (n,r,q)
   := {q\rm{VT}}_{a, c} (n,q) \cap S_{n,q} (r). \notag \\
  &{q\rm{SVT}}_{d, e, f} (n,r,q)
  := \bigl\{\bm{x} \in [q]^n
  \mid \textstyle\sum_{i=1}^{n} x_i \equiv f \pmod q , \\
  &\hspace{50mm} \sigma (\bm{x}) \in {\rm{SVT}}_{d, e}(n-1,r)\bigr\}, 
\end{align*}
where $a\in[n],c\in[q], d\in[r], e\in[2]$, and $f\in[q]$.
Schoney et al.\ \cite{schoeny2017novel} showed the following lemma:
\begin{lemma}[ {\cite[Lemma 1]{schoeny2017novel}} ]
 For all $d\in[r], e\in[2]$, and $f\in[q]$, the code ${q\rm{SVT}}_{d, e, f} (n,r,q)$ corrects a single insertion/deletion for a given $r-1$ adjacent positions.
\end{lemma}
As the result, they constructed the following non-binary single $b$-burst insertion/deletion correcting code:
\begin{align}
  \breve{C}_{q,b} 
  := 
  \{\bm{x} &\mid A_b(\bm{x})_1 \in {\rm{RLL\mathchar`-}}q{\rm{VT}}_{a,b}(n/b,r,q), \notag \\ 
  &~~\forall i \in [2,b]~~ A_b(\bm{x})_i \in {q\rm{SVT}}_{d,e,f}(n/b,r+2,q) \}. \label{eq:code_nonsch}
\end{align}

\section{Main Results}
This section constructs a non-binary burst insertion/deletion correcting code with a large cardinality.
Section \ref{ssec:cons} gives the main theorem and construction of the code.
Section \ref{ssec:proo} proves that the code is a non-binary burst insertion/deletion correcting code.
Section \ref{ssec:dec} provides the decoding algorithm for the code.
Section \ref{sec:card} will evaluate the asymptotic cardinality of the code and show a numerical example.

\subsection{Code Construction And Main Theorem \label{ssec:cons}}
We investigate the correcting capability of the non-binary SVT code.
As a result, we obtain that the code corrects a single insertion/deletion in a longer range as the following theorem.
\begin{theorem} \label{the:2}
 For all $d\in[r], e\in[2]$, and $f\in[q]$, the code ${q\rm{SVT}}_{d, e, f} (n,r,q)$ corrects a single insertion/deletion for a given $r$ adjacent positions.
\end{theorem}
Based on this result, we construct a code:
\begin{align}
  C_{q,b} 
  := 
  \{\bm{x} &\mid A_b(\bm{x})_1 \in {\rm{RLL\mathchar`-}}q{\rm{VT}}_{a,b}(n/b,r,q), \notag \\ 
  &~~\forall i \in [2,b]~~ A_b(\bm{x})_i \in {q\rm{SVT}}_{d,e,f}(n/b,r+1,q) \}. \label{eq:code_prop}
\end{align}
Moreover, we show the following theorem.
\begin{theorem} \label{the:3}
 The code $C_{q,b}$ corrects a single $b$-burst insertion/deletion.
\end{theorem}

\subsection{Proof of Theorems \label{ssec:proo}}
In this section, we prove Theorem \ref{the:2} and \ref{the:3}.
Now, we will derive several lemmas to prove Theorem \ref{the:2}
The following lemma clarifies the effect of a single deletion in a sequence to its ascent sequence.
\begin{lemma} \label{lem:1}
 Denote ${\bs u} = \sigma ({\bs x})$. 
 Then, $\sigma ({\bs x}_{\neg i}) = {\bs u}_{\neg (i-1)}$ or $\sigma ({\bs x}_{\neg i}) = {\bs u}_{\neg i}$ holds.
\end{lemma}
\begin{IEEEproof}
Denote ${\bs w} = \sigma ({\bs x}_{\neg i})$.
Obviously, it hold that $w_j = u_j$ for $j\in[1,i-2]$ and $w_j =u_{j+1}$ for $j\in[i,n-2]$.
Hence, we will show that $w_{i-1} = u_{i-1}$ or $w_{i-1} = u_{i}$ holds.

Firstly, we assume $x_{i-1}< x_{i} < x_{i+1}$.
Then, $u_{i-1} = u_{i} = 1$ holds.
Since $x_{i-1} < x_{i+1}$, $w_{i-1} = 1$ holds.
Hence, $w_{i-1} = u_{i-1} = u_{i} = 1$ holds.
Secondly, we assume $x_{i-1} < x_i$ and $x_i \ge x_{i+1}$.
Then, $u_{i-1} = 1$ and $u_{i} = 0$ holds.
If $x_{i-1} < x_{i+1}$, $w_{i-1}$ equals 1, otherwise $w_{i-1}$ equals 0.
Hence, $w_{i-1} = u_{i-1} = 1$ or $w_{i-1} = u_{i} = 0$ holds.

The other cases are proved in a similar way.
\end{IEEEproof}

Similarly, for an insertion, we obtain the following lemma.
\begin{lemma} \label{lem:posi_ins}
 Denote ${\bs u} = \sigma ({\bs x})$. 
 Then, $\sigma ({\bs x}_{\vdash (i, \lambda)}) = {\bs u}_{\vdash (i-1,\delta)}$ or $\sigma ({\bs x}_{\vdash (i,\lambda)}) = {\bs u}_{\vdash (i,\delta)}$ holds, where $\delta$ equals $0$ or $1$.
\end{lemma}

The following lemma is used for the proof of Theorem \ref{the:2}.
\begin{lemma} \label{lem:2}
 Consider ${\bs x}, {\bs y} \in \{ {\bs z} \in [q]^n \mid \sum_{i=1}^n z_i \equiv f \pmod{q} \}$ such that
 ${\bs x} \neq {\bs y}$ and ${\bs x}_{\neg s} = {\bs y}_{\neg t}$ for a pair of integers $s < t$.
 Denote ${\bs u} = \sigma ({\bs x})$, ${\bs v} = \sigma ({\bs y})$, and ${\bs w} = \sigma ({\bs x}_{\neg s})  = \sigma ({\bs y}_{\neg t})$.
 Then, the following hold:
 \begin{enumerate}
   \item If ${\bs w} = {\bs u}_{\neg (s-1)} = {\bs v}_{\neg t}$, 
    then there exist $i,j \in[s,t]$ such that $u_i \neq u_j$
   \item For a pair of integers $(\alpha,\beta) \in \{(0,0),(0,1),(1,1)\}$, 
     if ${\bs w} = {\bs u}_{\neg (s-\alpha)} = {\bs v}_{\neg (t-\beta)}$ and $u_{s-\alpha} = v_{t-\beta} = \gamma$,
     there exist $i \in [s-\alpha+1,t-\beta]$ such that $u_i \neq \gamma$.
 \end{enumerate}
\end{lemma}
\begin{IEEEproof}
 From Lemma \ref{lem:1}, we have ${\bs w} = {\bs u}_{\neg (s-1)}$ or ${\bs w} = {\bs u}_{\neg s}$, 
 and ${\bs w} = {\bs v}_{\neg (t-1)}$ or ${\bs w} = {\bs v}_{\neg t}$.
 Hence, ${\bs w} = {\bs u}_{\neg(s-\alpha)}={\bs v}_{\neg(t-\beta)}$ holds for a pair of integers $(\alpha,\beta) \in \{(0,0),(0,1),(1,0),(1,1)\}$.
 We have
 \begin{align*}
   0 
   &\equiv 
     {\textstyle \sum_{i=1}^n} x_i - {\textstyle \sum_{i=1}^n} y_i \pmod{q} \\
   &=
     x_s - y_t,
 \end{align*}
 where the first equivalence follows from ${\bs x}, {\bs y} \in \{ {\bs z} \in [q]^n \mid \sum_{i=1}^n z_i \equiv f \pmod{q} \}$
 and the second equation follows from ${\bs x}_{\neg s} = {\bs y}_{\neg t}$.
 Since $x_s, y_t \in[q]$, we get
 \begin{equation}
   x_s = y_t.     \label{eq:equiv_xy}
 \end{equation}
 From ${\bs x}_{\neg s} = {\bs y}_{\neg t}$ and ${\bs u}_{\neg (s-\alpha)} = {\bs v}_{\neg (t-\beta)}$, we have
 \begin{align}
    \label{eq:relation_xy}
    x_i &=
    \begin{cases}
      y_i, & ( i \in [1, s-1] \cup [t+1, n]),\\
      y_{i-1}, & ( i \in [s+1, t]),
    \end{cases} \\
    \label{eq:relation_uv1}
    u_i &=
    \begin{cases}
      v_i, & ( i \in [1, s-\alpha-1] \cup [t-\beta+1, n-1]),\\
      v_{i-1}, & ( i \in [s-\alpha+1, t-\beta]).
    \end{cases}
 \end{align}

 Firstly, we prove the case 1), i.e, the case of $(\alpha,\beta)=(1,0)$.
 Let us hypothesize $u_{s} = u_{s+1} = \cdots = u_t = 0$.
 From \eqref{eq:relation_uv1}, we get $v_{s-1} = v_{s} = \cdots = v_{t-1} = 0$.
 Hence, we have
 \begin{align}\label{eq:x_1_i}
   x_s \ge x_{s+1} \ge \cdots \ge x_{t+1}, \quad
   y_{s-1} \ge y_s \ge \cdots \ge y_t.
 \end{align}
 Note that $x_{t} = y_{t-1}$ and $x_{s+1} = y_s$ follow from \eqref{eq:relation_xy}.
 From \eqref{eq:equiv_xy}, \eqref{eq:relation_xy} and \eqref{eq:x_1_i}, we have
 \begin{align*}
   &x_s \ge x_{s+1} \ge \cdots \ge x_t = y_{t-1} \ge y_t = x_s, \\
   &x_s \ge x_{s+1} = y_s \ge y_{s+1} \ge \cdots \ge y_t = x_s.
 \end{align*}
 Note that both ends of these equations are $x_s$. Hence, these give
 \begin{align*}
   x_{s} = x_{s+1} = \cdots = x_{t} = y_s = y_{s+1} = \cdots = y_{t}.
 \end{align*}
 From this equation and \eqref{eq:relation_xy}, we get ${\bs x} = {\bs y}$.
 This contradicts ${\bs x} \neq {\bs y}$.
 Next, let us hypothesize $u_{s} = u_{s+1} = \cdots = u_{t} = 1$.
 Similarly, we get
 \begin{align*}
   x_s < x_{s+1} < \cdots < x_{t+1}, \quad
   y_{s-1} < y_s < \cdots < y_t.
 \end{align*}
 Note that $x_{s+1} = y_s$ follows from \eqref{eq:relation_xy}.
 Combining those and \eqref{eq:equiv_xy}, we have the following contradiction
 \begin{align*}
   x_s < x_{s+1} < \cdots < x_{t} = y_{t-1} < y_{t} = x_{s}.
 \end{align*}
 Thus, we obtain the case 1).

 Secondly, we prove the case 2), i.e, the case of $(\alpha,\beta)\in\{(0,0),(0,1),(1,1)\}$.
 From the assumption, we have $u_{s-\alpha} = v_{t-\beta} = \gamma$.
 Now, let us hypothesize $u_{i} = \gamma$ for all $i\in[s-\alpha+1,t-\beta]$.
 Suppose $\gamma = 0$.
 Then, $u_i = v_i = 0$ for all $i\in[s-\alpha,t-\beta]$.
 Hence, we have
 \begin{align}
   x_{s-\alpha} \ge x_{s-\alpha+1} \ge \cdots \ge x_{t-\beta+1}, \label{eq:x2ge} \\
   y_{s-\alpha} \ge y_{s-\alpha+1} \ge \cdots \ge y_{t-\beta+1}. \label{eq:y2ge}
 \end{align}
 Combining \eqref{eq:equiv_xy} \eqref{eq:relation_xy}, \eqref{eq:x2ge}, and \eqref{eq:y2ge}, we get
 \begin{alignat*}{7}
   &x_{s} & &x_{s} && && && && && \\
   &\rotatebox{90}{$=$}_{(\alpha=0)} & &\rotatebox{90}{$=$}_{(\alpha=1)} && && && && && \\   
   &x_{s-\alpha} &\ge& x_{s-\alpha+1} &\ge &x_{s-\alpha+2} &\ge& \cdots &\ge& x_{t-\beta} &\ge& x_{t-\beta+1} && \\
   & & &\rotatebox{90}{$=$}_{(\alpha=0)} && \rotatebox{90}{$=$} && && \rotatebox{90}{$=$} && \rotatebox{90}{$=$}_{(\beta=1)} && \\      
   && &y_{s-\alpha} &\ge &y_{s-\alpha+1} &\ge&  \cdots &\ge& y_{t-\beta-1} &\ge& y_{t-\beta} &\ge& y_{t-\beta+1} \\
   & & & && && &&  && \rotatebox{90}{$=$}_{(\beta=0)} && \rotatebox{90}{$=$}_{(\beta=1)} \\  
   & & & && && &&  && x_{s} && x_{s}
 \end{alignat*}
 where equality with label holds if the condition is satisfied (e.g, equality labeled with $(\alpha = 0)$ holds if $\alpha = 0$).
 The above gives
 \begin{equation*}
   x_s = x_{s+1} = \cdots = x_{t} =
   y_s = y_{s+1} = \cdots = y_{t},
 \end{equation*}
 for all pair of $(\alpha,\beta) \in \{(0,0),(0,1),(1,1)\}$.
 Combining this and \eqref{eq:relation_xy}, we get ${\bs x} = {\bs y}$.
 This contradicts ${\bs x} \neq {\bs y}$.
 Next, suppose $\gamma = 1$.
 Then, $u_i = v_i = 1$ for all $i\in[s-\alpha,t-\beta]$.
 Similarly, we get
 \begin{alignat*}{7}
   &x_{s} & &x_{s} && && && && && \\
   &\rotatebox{90}{$=$}_{(\alpha=0)} & &\rotatebox{90}{$=$}_{(\alpha=1)} && && && && && \\   
   &x_{s-\alpha} &<& x_{s-\alpha+1} &<& x_{s-\alpha+2} &<& \cdots &<& x_{t-\beta} &<& x_{t-\beta+1} && \\
   & & &\rotatebox{90}{$=$}_{(\alpha=0)} && \rotatebox{90}{$=$} && && \rotatebox{90}{$=$} && \rotatebox{90}{$=$}_{(\beta=1)} && \\      
   && &y_{s-\alpha} &<& y_{s-\alpha+1} &<&  \cdots &<& y_{t-\beta-1} &<& y_{t-\beta} &<& y_{t-\beta+1} \\
   & & & && && &&  && \rotatebox{90}{$=$}_{(\beta=0)} && \rotatebox{90}{$=$}_{(\beta=1)} \\  
   & & & && && &&  && x_{s} && x_{s}
 \end{alignat*}
 This leads the contradiction $x_s < x_s$.
 Thus, we obtain the case 2).
\end{IEEEproof}

Now we will prove the two theorems.

%% Begin proof of Theorem 1
{\it Proof of Theorem \ref{the:2}:}
Let us hypothesize that there exists a pair of codewords ${\bs x}, {\bs y}\in {q\rm{SVT}}_{d, e, f}(n,r,q)$ such that
${\bs x} \neq {\bs y}$ and ${\bs x}_{\neg s} = {\bs y}_{\neg t}$ for two integers $s<t$ and $t-s < r$.
Here, without loss of generality, we assume $s<t$.
Denote ${\bs u} = \sigma ({\bs x})$ and ${\bs v} = \sigma ({\bs y})$.
From Lemma \ref{lem:1}, 
$\sigma ({\bs x}_{\neg s}) = {\bs u}_{\neg (s-\alpha)}$ and $\sigma ({\bs y}_{\neg t}) = {\bs v}_{\neg (t-\beta)}$ holds for a pair of integers $(\alpha,\beta) \in \{(0,0),(0,1),(1,0), (1,1)\}$.
We have
\begin{align}
  0 
  &\equiv
  {\textstyle \sum_{i=1}^{n-1}} u_i - {\textstyle \sum_{i=1}^{n-1}} v_i
  \pmod 2 \notag
  \\
  &= 
  u_{s-\alpha} - v_{t-\beta}, \notag
\end{align}
where the first equivalence follows from ${\bs x},{\bs y} \in {q\rm{SVT}}_{d, e, f}(n,r,q)$, i.e, ${\bs u}, {\bs v} \in {\rm SVT}_{d,e}(n-1,r)$, 
and the second equation follows from ${\bs u}_{\neg (s-1)} = \sigma ({\bs x}_{\neg s}) = \sigma ({\bs y}_{\neg t}) = {\bs v}_{\neg t}$.
Hence, we get
\begin{equation}
 u_{s-\alpha} = v_{t-\beta}. \label{eq:uv_equiv}
\end{equation}
Since ${\bs u}_{\neg (s-\alpha)} = {\bs v}_{\neg (t-\beta)}$, we get \eqref{eq:relation_uv1}.
From \eqref{eq:relation_uv1} and \eqref{eq:uv_equiv}, we have
\begin{align} 
  &~{\textstyle \sum_{i=1}^{n-1} } iu_i  - {\textstyle \sum_{i=1}^{n-1}} i v_i
  \notag \\
  =&~
  {\textstyle \sum_{i=s-\alpha+1}^{t-\beta}} u_i
  -(t-s+\alpha-\beta)u_{s-\alpha}  .
  \label{eq:C_red}
\end{align}
Note that ${\bs x}, {\bs y} \in q{\rm SVT}_{d,e,f}(n,r,q) \subset \{ {\bs z}\in[q]^n \mid \sum_{i=1}^{n}z_i \equiv f \pmod{q}\}$.
Hence, the pair of ${\bs x}$ and ${\bs y}$ satisfies the conditions of Lemma \ref{lem:2}.
Firstly, we assume $(\alpha,\beta) = (1,0)$.
Then, case 1) of Lemma \ref{lem:2} derives 
\begin{align*}
  0 <
  {\textstyle \sum_{i=s}^{t}} u_i
  \le t-s. 
\end{align*}
Recall that $t-s < r$.
Combining the above with \eqref{eq:C_red}, we obtain for $u_{s-1} = 0$
\begin{align*}
 &0 <  
  {\textstyle \sum_{i=1}^{n-1} } iu_i  - {\textstyle \sum_{i=1}^{n-1}} i v_i
  \le t - s < r,
\end{align*}
and for $u_{s-1} = 1$
\begin{align*}
 &-r \le -(t-s)-1 <  
  {\textstyle \sum_{i=1}^{n-1} } iu_i  - {\textstyle \sum_{i=1}^{n-1}} i v_i
  \le -1. 
\end{align*}
However, these contradict ${\textstyle \sum_{i=1}^{n-1} } iu_i  - {\textstyle \sum_{i=1}^{n-1}} i v_i \equiv 0 \pmod{r}$ which follows from ${\bs x},{\bs y} \in q{\rm SVT}_{d, e, f}(n,r,q)$, i.e, ${\bs u}, {\bs v} \in {\rm SVT}_{d,e}(n-1,r)$.
Secondly, we assume $(\alpha,\beta)\in\{(0,0),(0,1),(1,1)\}$.
Then, case 2) of Lemma 4 derives
\begin{align*}
  &0 <
  {\textstyle \sum_{i=s-\alpha+1}^{t-\beta} } u_i
  \le t-s+\alpha-\beta, & &\text{(if $u_{s-\alpha} = 0$)}, \\
  &0 \le
  {\textstyle \sum_{i=s-\alpha+1}^{t-\beta} } u_i
  < t-s+\alpha-\beta, & &\text{(if $u_{s-\alpha} = 1$)} .
\end{align*}
Since $t-s < r$ and $\alpha-\beta \le 0$, we have $t-s + \alpha - \beta<r$
Combining the above and \eqref{eq:C_red}, we obtain
\begin{align*}
  &0 <
  {\textstyle \sum_{i=1}^{n-1} } iu_i  - {\textstyle \sum_{i=1}^{n-1}} i v_i  
  < r , & &\text{(if $u_{s-\alpha} = 0$)}, \\
  &-r <
      {\textstyle \sum_{i=1}^{n-1} } iu_i  - {\textstyle \sum_{i=1}^{n-1}} i v_i  
  < 0, & &\text{(if $u_{s-\alpha} = 1$)} .
\end{align*}
Similarly, these contradict ${\textstyle \sum_{i=1}^{n-1} } iu_i  - {\textstyle \sum_{i=1}^{n-1}} i v_i \equiv 0 \pmod{r}$.
Hence, we obtain the theorem.
\hfill\QED
%% End proof of Theorem 1

Theorem \ref{the:3} is proved in a similar way to \cite[Theorem 5]{schoeny2017codes}.

\subsection{Decoding Algorithm \label{ssec:dec}}
Due to space limitations, we only describe the insertion/deletion correcting algorithm for the non-binary SVT code.
In other words, we omit the decoding algorithm for $C_{q,b}$.

We denote the remainder when $i$ is divided by $q$, by $\langle i \rangle_q$.
Denote the transmitted sequence, by ${\bs x}$.
Algorithms \ref{algo:delcor} and \ref{algo:inscor} describe the deletion and insertion correcting algorithm for the SVT code, respectively.
The set of inputs of those algorithms is the received sequence ${\bs y}$, code parameters $(d,e,f,n,r)$, and first possible deletion/insertion position $k$.
The output of those algorithms is the estimated sequence.

In Algorithm \ref{algo:delcor}, $\hat{x}$ stands the deleted symbol and $j$ represents the position of the deleted symbol.
Step \ref{stp:del1} calculates the deleted symbol since $\sum_{i=1}^{n-1} y_i + \hat{x} = \sum_{i=1}^n x_i \equiv f \pmod{q}$.
Step \ref{stp:del2} checks whether the $k$-th symbol is deleted.
If the condition of Step \ref{stp:del2} does not satisfy, then the deletion position is in $[k+1,k+r-1]$.
In such a case, from Lemma \ref{lem:1}, $\sigma ({\bs y})$ equals to $\sigma ({\bs x})_{\neg i}$ with an integer $i \in [k,k+r-1]$.
Hence, we obtain ${\bs u} = \sigma ({\bs x})$ as in Step \ref{stp:del3}.
The algorithm searches the position of the deleted symbol in Steps \ref{stp:del4}-\ref{stp:del5}.

In Algorithm \ref{algo:inscor}, $\hat{x}$ stands the inserted symbol and $j$ represents the position of the inserted symbol.
Step \ref{stp:ins1} calculates the inserted symbol since $\sum_{i=1}^{n+1} y_i - \hat{x} = \sum_{i=1}^n x_i \equiv f \pmod{q}$.
Step \ref{stp:ins2} checks whether the $k$-th symbol is inserted.
If the condition of Step \ref{stp:ins2} does not satisfy, then the inserted position is in $[k+1,k+r]$.
In such case, from Lemma \ref{lem:posi_ins}, $\sigma ({\bs y})$ equals $\sigma ({\bs x})_{\vdash (i,\delta)}$ with an integer $i \in[k,k+r]$ and $\delta \in \{0,1\}$.
Hence, we obtain ${\bs u} = \sigma ({\bs x})$ as in Step \ref{stp:ins3}.
The algorithm searches the position of the inserted symbol in Steps \ref{stp:ins4}-\ref{stp:ins5}.

\begin{algorithm}[tb]
  \caption{Deletion Correction for $q{\rm SVT}_{d, e, f}(n, r, q)$ \label{algo:delcor}}
  \begin{algorithmic}[1]
    \REQUIRE Received sequence ${\bs y}$, code parameters $(d,e,f,n,r)$, first possible deletion position $k$
    \ENSURE Estimated sequence $\bm{x}'$ %(equal to original sequence $\bm{x}$)
    \STATE \label{stp:del1} $\hat{x} \leftarrow \bigl\langle f - \sum_{i=1}^{n-1} y_i \bigr\rangle_q$
    \IF{${\bs y}_{\vdash (k,\hat{x})} \in q{\rm SVT}_{d, e, f}(n, r, q)$} \label{stp:del2}
    \STATE $\bm{x}' \gets {\bs y}_{\vdash (k,\hat{x})}$
    \ELSE
    \STATE $(\bm{u}, [s, t]) \gets {{\sf SVT{\mathchar`-}DC}}(\sigma({\bs y}), d, e, n-1,r, k)$ \label{stp:del3}
    \STATE $s' \gets \max\{s, k+1\}$, $t' \gets \min\{t, k+r-2\}$
    \STATE $j \gets t'+1$ \label{stp:del4}
    \IF{$u_{s'} = 0$ (i.e, $u_{s'} = u_{s'+1} = \cdots = u_{t'} = 0$)}
    \FOR{$i = s', s'+1, \dots, t'$}
    \IF{$\hat{x} \ge y_i$}
    \STATE $j \gets i$ and go to Step \ref{stp:del6}
    \ENDIF
    \ENDFOR
    \ELSE 
    \FOR{$i= s', s'+1, \dots, t'$}
    \IF{$\hat{x} < y_i$}
    \STATE $j \gets i$ and go to Step \ref{stp:del6}
    \ENDIF
    \ENDFOR
    \ENDIF \label{stp:del5}
    \STATE $\bm{x}' \gets {\bs y}_{\vdash (j,\hat{x})}$ \label{stp:del6}
    \ENDIF 
  \end{algorithmic}
\end{algorithm}

\begin{algorithm}[tb]
  \caption{Insertion Correction for $q{\rm SVT}_{d, e, f}(n, r, q)$ \label{algo:inscor}}
  \begin{algorithmic}[1]
    \REQUIRE Received sequence ${\bs y}$, code parameters $(d,e,f,n,r)$,
    first possible insertion position $k$
    \ENSURE Estimated sequence $\bm{x}'$ %(equal to original vector $\bm{x})$
    \STATE \label{stp:ins1} $\hat{x} \gets \bigl\langle \sum_{i=1}^{n+1} y_i - f \bigr\rangle_q$
    \IF{${\bs y}_{\neg k} \in q{\rm SVT}_{d, e, f}(n, r, q)$} \label{stp:ins2}
    \STATE ${\bs x}' \gets {\bs y}_{\neg k}$
    \ELSE
    \STATE $(\bm{u}, [s, t]) \gets {\sf{SVT \mathchar`- IC}}(\sigma({\bs y}), d, e, n-1,r, k)$ \label{stp:ins3}
    \STATE $s' \gets \max\{s,k+1\}$, $t' \gets \min\{t, k+r-1\}$
    \FOR{$i = s',s'+1,\dots, t'+1$} \label{stp:ins4}
    \IF{$y_i = \hat{x}$}
    \STATE $j \leftarrow i$ and go to Step \ref{stp:ins6}
    \ENDIF
    \ENDFOR \label{stp:ins5}
    \STATE ${\bs x}' \gets {\bs y}_{\neg j}$ \label{stp:ins6}
    \ENDIF 
  \end{algorithmic}
\end{algorithm}

\section{The Number of Codewords \label{sec:card} }
This section evaluates the gap between the lower bound of the cardinality of the constructed code and the upper bound of the cardinality of arbitrary non-binary $b$-burst insertion/deletion correcting codes.
Moreover, we evaluates the number of codewords of the SVT codes by a numerical example for an evidence that the code in \eqref{eq:code_prop} has a larger cardinality.

\subsection{Lower Bound of Cardinality of Constructed Code}
In a similar way to \cite[Lemma 2]{schoeny2017codes}, we have the following lemma.
\begin{lemma}
 The following holds
  \begin{align*}
    | S_{n,q} (r)| \ge (q^r - n) q^{n-r}.
  \end{align*}
\end{lemma}

By the pigeonhole principle and this lemma, we get the following two lemmas.
\begin{lemma}
  The cardinality of non-binary run-length-limited VT code is lower bounds as:
  \begin{align*}
    \max_{ a \in [n], c \in [q]} |{\rm{RLL\mathchar`-}}q{\rm{VT}}_{a, c} (n,r,q)|
    \ge
    \frac {(q^r - n) q^{n-r-1}} {n}.
  \end{align*}
\end{lemma}
\begin{lemma}
  The cardinality of non-binary SVT code is lower bounds as:
  \begin{align*}
    \max_{ d \in [r], e \in [2], f \in [q]} |{q\rm{SVT}}_{d, e, f}(n,r,q)|
    \ge
    \frac {q^{n-1}} {2r}.
  \end{align*}
\end{lemma}
From those lemmas, we obtain a lower bound of cardinality of the constructed code.
\begin{theorem}
  For all $r$, the cardinality of $C_{q,b}$ satisfies
  \begin{align}
    \max |C_{q,b}| 
    \ge
    \frac{q^{n-b}}{n}
    \frac {b - n q^{-r}} {2^{b-1}(r+1)^{b-1}}.
    \label{eq:Cq_Size}
  \end{align}
\end{theorem}
Substituting $r = \log_q n$ in \eqref{eq:Cq_Size}, we have
\begin{align}
   \max|C_{q,b}|
  &\ge 
    \frac {2 q^{n-b}} {n} \cdot \frac {b-1} {2^b (\log_q n+1)^{b-1}}. \label{eq:Cq_Up}
\end{align}

We define {\it redundancy} of a $q$-ary code $C$ by $n- \log_q|C|$.
From \eqref{eq:Cq_Up}, an upper bound of redundancy of $C_{q,b}$ with the best parameter is
\begin{align}
  &b + \log_q n - \log_q (b-1) + (b-1)\log_q 2 \notag \\
  &+ (b-1)\log_q (\log_q n+1). \label{eq:redu_Cq}
\end{align}

\subsection{Upper Bound of Cardinality of Burst Insertion/Deletion Correcting Code}
Let $\mathcal{C}$ be the set of non-binary $b$-burst insertion/deletion correcting codes of length $n$.
Define $M_{b}(n) := \argmax_{C\in\mathcal{C}}|C|$. 
In words, $M_b(n)$ is the non-binary $b$-burst insertion/deletion correcting code of length $n$ with maximum cardinality, i.e, $M_b(n)$ is the best code.
The following theorem gives an upper bound of the cardinality of $M_b(n)$.
\begin{theorem} \label{the:5}
 For enough large $n$, the following holds:
 \begin{align*}
  |M_{b}(n)| \le \frac{q^{n-b+1}}{(q-1)n}.
 \end{align*}
\end{theorem}
This theorem is proved in a similar way to \cite[Lemma 1]{levenshtein1967asymptotically}.
\begin{IEEEproof}
  Define $m := n/b -1$.
  Denote the number of runs in ${\bs x}$, by $||{\bs x}||$.
For a positive integer $r$, define
\begin{align*}
  M_1(r)
  &:=
  \{ {\bs x} \in M_b(n) \mid \forall i \in [1,b]~  ||A_b({\bs x})_{i}|| \ge r+2\}, \\
  M_2(r)
  &:=
  \{ {\bs x} \in M_b(n) \mid \exists i \in [1,b] \text{~s.t.~} ||A_b({\bs x})_{i}|| \leq r+1\}.
\end{align*}
Note that $M_b(n) = M_1(r) \cup M_2(r)$ and $M_1(r) \cap M_2(r) = \emptyset$.
This leads
\begin{equation}
  |M_b(n)| = |M_1(r)| + |M_2(r)| \quad \text{for all $r$}.
  \label{eq:mb}
\end{equation}

Now, we will derive upper bounds of $|M_1(r)|$ and $|M_2(r)|$.
Firstly, we consider $|M_1(r)|$.
Denote $D({\bs x}) := \{ {\bs x}_{\neg [i,i+b-1]} \mid i\in[1,n-b+1] \}$.
In words, $D({\bs x})$ is $b$-burst deletion ball for ${\bs x}$, i.e, the set of sequences after $b$-burst deletion to ${\bs x}$.
The volume of $b$-burst deletion ball for ${\bs x}$ is derived in \cite{levenshtein1967asymptotically} as
\begin{align*}
  |D({\bs x})| 
  &= 
  1 + \textstyle\sum_{i=1}^b (|| A_b({\bs x})_{i}||-1).
\end{align*}
Since $|| A_b({\bs x})_{i}|| \ge r+2$ for all $i$, $|D({\bs x})|$ is bounded by
\begin{align*}
  |D({\bs x})|
  \ge
  b(r+1) + 1.
\end{align*}
Since $M_b(n)$ is a $b$-burst deletion correcting code, 
$M_1(r)$ is also a $b$-burst deletion correcting code.
Hence, $\bigcup_{{\bs x}\in M_1(r)} D({\bs x}) \subseteq [q]^{n-b}$ and $D({\bs x}) \cap D({\bs y})$ for all ${\bs x}, {\bs y} \in M_{1}(r)$ hold.
This leads $q^{n-b} \ge \sum_{{\bs x}\in M_1(r)} |D({\bs x})|$.
Combining the above yields
\begin{align*}
  q^{n-b}
  \ge 
  |M_1(r)| \{b(r+1) + 1\} .
\end{align*}
As the result, we have an upper bound for $|M_1(r)|$ as follows:
\begin{equation}
  |M_1(r)| \le \frac{q^{n-b}}{b(r+1) + 1}
  < \frac{q^{n-b}}{b(r+1)}=: f(r).
  \label{eq:up_m1}
\end{equation}
Secondly, we derive an upper bound for $|M_2(r)|$.
Define
\begin{align*}
  B_{\le r+1 }
  &:= 
  \{{\bs x} \in [q]^n \mid
  \exists i \text{~s.t.~} ||A_b({\bs x})_{i}|| \leq r+1\}, \\
  B_{\le r+1, i}
  &:=
  \{{\bs x} \in [q]^n \mid ||A_b({\bs x})_{i}|| \leq r+1\}, \\
  B_{j,i}
  &:=
  \{{\bs x} \in [q]^n \mid ||A_b({\bs x})_{i}|| = j\}.
\end{align*}
Then, the following holds
\begin{equation*}
  M_2(r)
  \subseteq
  B_{\le r+1}
  =
  {\textstyle \bigcup_{i=1}^{b}} B_{\le r+1, i}
  =
  {\textstyle \bigcup_{i=1}^{b}\bigcup_{j=1}^{r+1}} B_{j, i}. 
\end{equation*}
Now, the cardinality of $B_{j,i}$ is
\begin{equation*}
  |B_{j,i}|
  =
  \binom{m}{j-1}q^{n-m}(q-1)^{j-1}.
\end{equation*}
These derives 
\begin{align*}
  |M_2(r)|
  \le
  b q^{n-m} \sum_{j=0}^{r} \binom{m}{j}(q-1)^{j}.
\end{align*}
For $r < (1-q^{-1})m$, the summation is bounded by (e.g, see \cite[Exercise 5.8]{gallager1968information})
\begin{equation}
  \sum_{j=0}^{r} \binom{m}{j}(q-1)^{j}
  \le
  (q-1)^r \exp [ m h_2(r/m) ],
\end{equation}
where $h_2(x) := -x\ln x -(1-x)\ln (1-x)$.
For $r \ge \frac{q-1}{q}m$, 
\begin{align*}
  \sum_{j=0}^{r} \binom{m}{j}(q-1)^{j}
  &\le
  q^m - (q-1)^{r+1} \frac{\exp[ m h_2((r+1)/m) ]}{\sqrt{2m}} 
\end{align*}
where the last inequality follows from $\sum_{j=r+1}^m\binom{m}{j} \ge \binom{m}{r+1}(q-1)^{r+1}$ and $\binom{m}{r+1} \ge \exp[ m h_2((r+1)/m) ] / \sqrt{2m}$.
Thus, $|M_2(r)|$ is bounded by
\begin{align}
  |M_2(r)|
  &\le g(r) :=
  \begin{cases}
    g_1(r), & \text{if $r < \frac{q-1}{q}m$},   \\
    g_2(r), & \text{if $r \ge \frac{q-1}{q}m$},
  \end{cases}
  \label{eq:up_m2} \\
  g_1(r) &:= bq^{n-m} (q-1)^r \exp [ m h_2(r/m) ], \notag \\
  g_2(r) &:= bq^{n-m} \left[  q^m - (q-1)^{r+1} \frac{\exp[ m h_2((r+1)/m) ]}{\sqrt  {2m}} \right].
  \notag 
\end{align}

Combining \eqref{eq:mb}, \eqref{eq:up_m1} and \eqref{eq:up_m2} yields
\begin{align}
  |M_b(n)|
  \le
  \min_{r}
  \bigr( f(r) + g(r) \bigl).
  \label{eq:mb_up1}
\end{align}
Note that $f(r)$ is monotonically decreasing function and $g_1(r),g_2(r)$ are monotonically increasing functions.
Firstly, we consider the case of $r < \frac{1-q}{q}m$.
Let $\alpha$ be a positive real number.
Define $\epsilon := \sqrt{\frac{\alpha \ln m}{m}}$.
Substituting $r = (1-q^{-1} -\epsilon)m $ yields
\begin{align*}
  f\Bigl( \bigr(1-q^{-1}-\epsilon\bigr) m \Bigr)
  &\le
  \frac{q^{n-b}}{b}
  \frac{1}{\frac{q-1}{q}(m+1) - \epsilon m}
  \\ &=
  \frac{q^{n-b+1}}{n(q-1)}
  \left( 1 + O(\epsilon) \right),
\end{align*}
where the last equation follows from $(1-\epsilon)^{-1} = 1 + O(\epsilon)$.
Note that $\ln (1+x) = x - \frac{1}{2}x^2 + O(x^3) $.
This leads
\begin{align*}
  h_2\bigl( 1-q^{-1}-\epsilon   \bigr)
   &=
  -\left( 1-q^{-1} - \epsilon  \right)  \ln (q-1)
  + \ln q \\
  &~~~~- \frac{1}{2} \frac{q^2}{q-1}\epsilon^2
    +O(\epsilon^3)
\end{align*}
This yields
\begin{align*}
  g_1\Bigl( \bigl(1-q^{-1} -\epsilon \bigr)m  \Bigr)
  =
  bq^{n} m^{ -\frac{1}{2} \frac{q^2}{q-1}\alpha }
  \exp \left[  O(\epsilon^3 m) \right].
\end{align*}
Hence, if $\alpha > 2\frac{q-1}{q^2}$, $g_1(r) = o(f(r))$.
Otherwise, $f(r) = O(g_1(r))$.
Thus, for $r<\frac{q-1}{q}m$, \eqref{eq:mb_up1} is evaluated as
\begin{align}
  \min_{r<\frac{q-1}{q}m} (f(r)+ g(r))
  &\le
  f\left( (1-q^{-1})m -\sqrt{m\ln m} \right) \notag
  \\ &=
  \frac{q^{n-b+1}}{n(q-1)}
  \biggl(1+O\biggl(\sqrt{\frac{\ln m}{m}} \biggr)\biggr). \label{eq:mb_up2}
\end{align}
Secondly, let us consider the case of $r\ge \frac{1-q}{q}m$.
Recall that $f(r)$ and $g_2(r)$ are monotonically decreasing and increasing function, respectively.
Since $f((1-q^{-1})m) < g_2((1-q^{-1})m)$, $f(r)<g_2(r)$ holds for all $r\ge \frac{q-1}{q}m$.
Thus, for $r \ge \frac{q-1}{q}m$, \eqref{eq:mb_up1} is evaluated as
\begin{equation}
  \min_{r\ge\frac{q-1}{q}m} (f(r)+ g_2(r))
  =
  bq^n + o(q^n). \label{eq:mb_up3}
\end{equation}
Comparing \eqref{eq:mb_up2} and \eqref{eq:mb_up3} leads the theorem.
\end{IEEEproof}

From Theorem \ref{the:5}, the redundancy of $M_b(n)$ is lower bounded by
\begin{align*}
 b - \log_q 2 + \log_q n.
\end{align*}
By comparing \eqref{eq:redu_Cq}, the gap of redundancy between the constructed code and the best code is upper bounded by
\begin{align*}
 - \log_q (b-1) + b \log_q 2 + (b-1)\log_q (\log_q n+1).
\end{align*}

\subsection{Numerical Example}
\begin{table}[tb]
  \centering
  \caption{The cardinality of non-binary SVT codes with best parameters for $n=10,q=4$.   \label{tab:1}}
  \begin{tabular}{|c|| c|c|c|c|c|} \hline
    $r$ & 2 & 3 & 4 & 5 & 6 \\ \hline
    Cardinality &
    66240 & 44028 & 33136 & 26475 & 22108 
    \\ \hline \hline
    $r$ & 7 & 8 & 9 & 10 & \\ \hline
    Cardinality & 19000 & 17874 & 17918 & 18156 & \\ \hline
  \end{tabular}
\end{table}

Table \ref{tab:1} shows the number of codewords of the non-binary SVT code with best parameters for $n=10,q=4$, i.e, shows $\max_{d,e,f}|q\mathrm{SVT}_{d,e,f}(10,r,4)|$ for $r=2,3,\dots,10$.
From Table \ref{tab:1}, we see that the number of codewords decreases for $r\le 8$ as $r$ increases.
In other $n,q$, we also observe the number of codewords decreases except that $r$ is nearly equals to $n$.
Hence, for small $r$ (e.g, $r=\log_q n$, employed in \eqref{eq:Cq_Up}), we conclude that $C_{q,b}$ has a larger cardinality than $\breve{C}_{q,b}$.

\section{Conclusion and Future Works}
In this paper, we have constructed a non-binary $b$-burst insertion/deletion correcting code with a larger cardinality and presented a decoding algorithm for the code.
We also have derived a lower bound on the cardinality of the proposed code and an asymptotic upper bound on the cardinality of non-binary $b$ burst deletion correcting codes.
Our future works are (1) construction of non-binary codes which correct a deletion burst of {\sl at most} $b$ consecutive symbols and (2) deriving non-asymptotic upper bound on the maximum cardinality of any non-binary $b$ burst deletion correcting code.

\section*{Acknowledgment}
The authors wish to thank to Dr.~C.~Schoeny for telling us \cite{schoeny2017novel}.
This work was supported by JSPS KAKENHI Grant Number 16K16007.

\end{document}